# Purcell enhanced and blinking free single photons from InAs/GaAs quantum dots in deterministically placed circular Bragg gratings

Authors: Peter Gschwandtner*[1], Quirin Buchinger*[1], Krishna Chand Maurya[1], Mohamed Helal[1], Barbara Souza Damasceno[1], Ravindra Kumar[1], Hyemin Kim[1,2], Ievgen Brytavskyi[3], Silke Kuhn[1], Arne Ludwig[4], Dirk Reuter[5], Yong-Hoon Cho,[2] Tobias Huber-Loyola[1,6], Sven Höfling[1]

[1]Julius-Maximilians-Universität Würzburg, Faculty of physics and astronomy, Technische Physik, Würzburg, Germany

[2]Korea Advanced Institute of Science and Technology (KAIST), Department of Physics, Daejeon, Republic of Korea

[3]Johannes Kepler Universität Linz, Institute of Semiconductor and Solid State Physics, Semiconductor Physics Division, Linz, Austria

[4]Ruhr-Universität Bochum, Lehrstuhl für Angewandte Festkörperphysik, Bochum, Germany

[5]Paderborn University, Department of Physics, Paderborn, Germany

[6]Karlsruhe Institute of Technology, Institute of Photonics and Quantum Electronics (IPQ) & Center for Integrated Quantum Science and Technology (IQST), Karlsruhe, Germany

**Corresponding author:**

Peter Gschwandtner peter.gschwandtner@uni-wuerzburg.de

Sven Höfling sven.hoefling@uni-wuerzburg.de

## Abstract

The development of efficient, deterministic, and tunable single-photon sources is a cornerstone for the realization of long-distance quantum communication, quantum repeaters, and photonic quantum computing technologies. In this study, we demonstrate a bright, charge-tunable single-photon source in the 900 nm wavelength range based on InAs quantum dots (QDs) embedded in a p-i-n doped GaAs membrane, which shows blinking free emission.

We use a modified circular Bragg grating (CBG) as a micro-resonator. By adding fourfold symmetric bridges in a labyrinth-like geometry, we provide a conductive pathway to the central disk, thereby enabling electrical contact to the QD while maintaining high Purcell enhancement and photon extraction efficiency (PEE). For the negative trion ($X^-$), we demonstrate a lifetime of 44.3 ± 0.2 ps - corresponding to a Purcell factor of 18.0 ± 0.7 and a PEE of 68.1% ± 3.1%. Furthermore, the device is blinking-free with very low multi-photon contribution, evidenced by a second-order autocorrelation value $g^{(2)}(0) < 0.017 \pm 0.015$. By applying a vertical diode bias, we demonstrate precise charge-state control, resolving distinct emission plateaus ranging from the single negatively charged trion ($X^-$) to triply negatively charged excitons ($X^{3-}$).

These results showcase a robust architecture that simultaneously provides high efficiency, high repetition rates, and deterministic charge control, fulfilling key requirements for the next generation of quantum network hardware.

**Keywords**: *quantum dots, circular Bragg grating, bullseye cavity, single-photon source, charge tuning, Purcell enhanced emission, blinking free*

## Introduction

The realization of a global quantum communication network, hinges on the reliable distribution of entanglement over long distances [1]. However, the transmission of photonic qubits through optical fibers is limited by absorption and decoherence, both of which grow exponentially with distance. Quantum repeaters offer a solution to these limitations [2], enabling the generation of entangled photon pairs across arbitrarily large distances. A central prerequisite for realizing these devices is the development of efficient, deterministic, and tunable entangled-photon sources (EPS). Semiconductor quantum dots (QDs) have emerged as a promising platform for on-demand entangled-photon generation [3]. They offer high brightness, low-multiphoton contributions and high indistinguishability [4], as well as compatibility with mature semiconductor nanofabrication techniques.

Extensive work has focused on growing QDs that directly emit at 1550 nm [5,6]. However, meeting the stringent performance demands - such as simultaneous high brightness, indistinguishability, and low-multiphoton contributions - remains an open challenge. An alternative approach involves quantum frequency down-conversion (QFDC) of near-infrared QDs to the C-band [7,8]. This strategy leverages the mature growth and nanofabrication techniques established for 900 nm semiconductor QDs at the cost of some extra loss that comes with the QFDC.

To overcome the limited photon extraction efficiencies of planar semiconductor membranes, QDs can be integrated into microlenses [9] and waveguides [10,11] or into optical microcavities such as micropillars [12] and circular Bragg gratings (CBGs) [13]. Among these optical cavities, CBGs have attracted significant interest [14–18], because they offer a high collection efficiency over a broad wavelength range as well as high Purcell enhancement. The latter is reducing the exciton lifetime and allows for increasing the maximal repetition rate [19]. Furthermore, carefully engineered CBG modes can enhance multiple transitions from the same QD over a range of several nanometers [15,20]. This relatively broad cavity mode up to ~10 nm allows for wavelength shifts, e.g. via Stark tuning in an external electric field or through strain tuning [21], with only a small reduction in Purcell enhancement due to spectral mismatch.

Another desirable feature of single-photon sources is external control over the QD charge states. Certain quantum communication protocols rely on specific charge configurations [22], which can be achieved by embedding the QD in field-effect structures such as a p-i-n diode [23,24] . Such a charge-tunable device has the additional benefit of suppressing blinking [25], i.e., the random fluctuation in the charges captured by the QD. Blinking limits the quantum efficiency of the device by introducing random dead-times in the photon stream. Indeed, a blinking free single photon source is a prerequisite for many applications where a steady stream of photons is a key requirement, for example quantum cryptography or cluster state generation.

However, these features are not always mutually compatible. For example, a typical CBG design consists of a central disk surrounded by fully etched trenches. This geometry would electrically isolate the central pillar, negating charge control. This challenge can be overcome by introducing labyrinth-like bridges between the sequential rings [26]. Further theoretical calculations have demonstrated that these bridges do not negatively impact the device performance, if the design parameters are chosen carefully, with theoretically PEE values in excess of 90 % [27]

Consequently, realizing a bright, charge-tunable single-photon source requires careful consideration of several factors and design parameters, including the epitaxial layer structure, material refractive indices, and the overall geometry of the sample. In this study, we report on realizing a single-photon source with electrically contacted CBG with the labyrinth design as proposed by Buchinger et al [26].

## Sample Design and Growth

Our sample design utilizes epitaxially grown self-assembled InAs QDs embedded in a half-wavelength (λ/2) GaAs membrane, optimized for emission in the 920 nm spectral range. The GaAs membrane incorporates a p-i-n diode structure to enable deterministic charge state tuning. An additional $Al_{0.85}Ga_{0.15}As$ barrier is integrated between the QDs and the p-doped region of the membrane to suppress hole tunneling between the QD and the p-type contact.

To achieve high photon extraction efficiency (PEE), the design includes a highly reflective backside mirror composed of an $Al_2O_3$ spacer layer and a gold mirror. FDTD simulations of the CBG design based on Ref. [26], extended to include the backside mirror, predict a PEE of 58% into an objective with NA=0.64 alongside with a Purcel factor of 22.

The electronic band structure of the device was modeled using the *nextnano* simulation software. Figure 1 illustrates the conduction band under zero bias voltage (dashed lines) and applied bias $V_b$ (solid lines). At zero bias, the built-in potential of the p-i-n diode tilts the energy bands, positioning the lowest energy states in the QD above the quasi-Fermi Level of the n-type contact. By applying forward bias voltage $V_b$ the energy bands are flattened until the QD electronic states reach resonance with the Fermi level of the electron reservoir on the n-doped region. This enables controlled electron tunneling between the electron reservoir and the QD. At the same time, the $Al_{0.85}Ga_{0.15}As$ barrier suppresses tunneling between QD hole energy states and the quasi-Fermi Level on p-type contact. This allows for selective charge control with a focus on negatively charged trions.

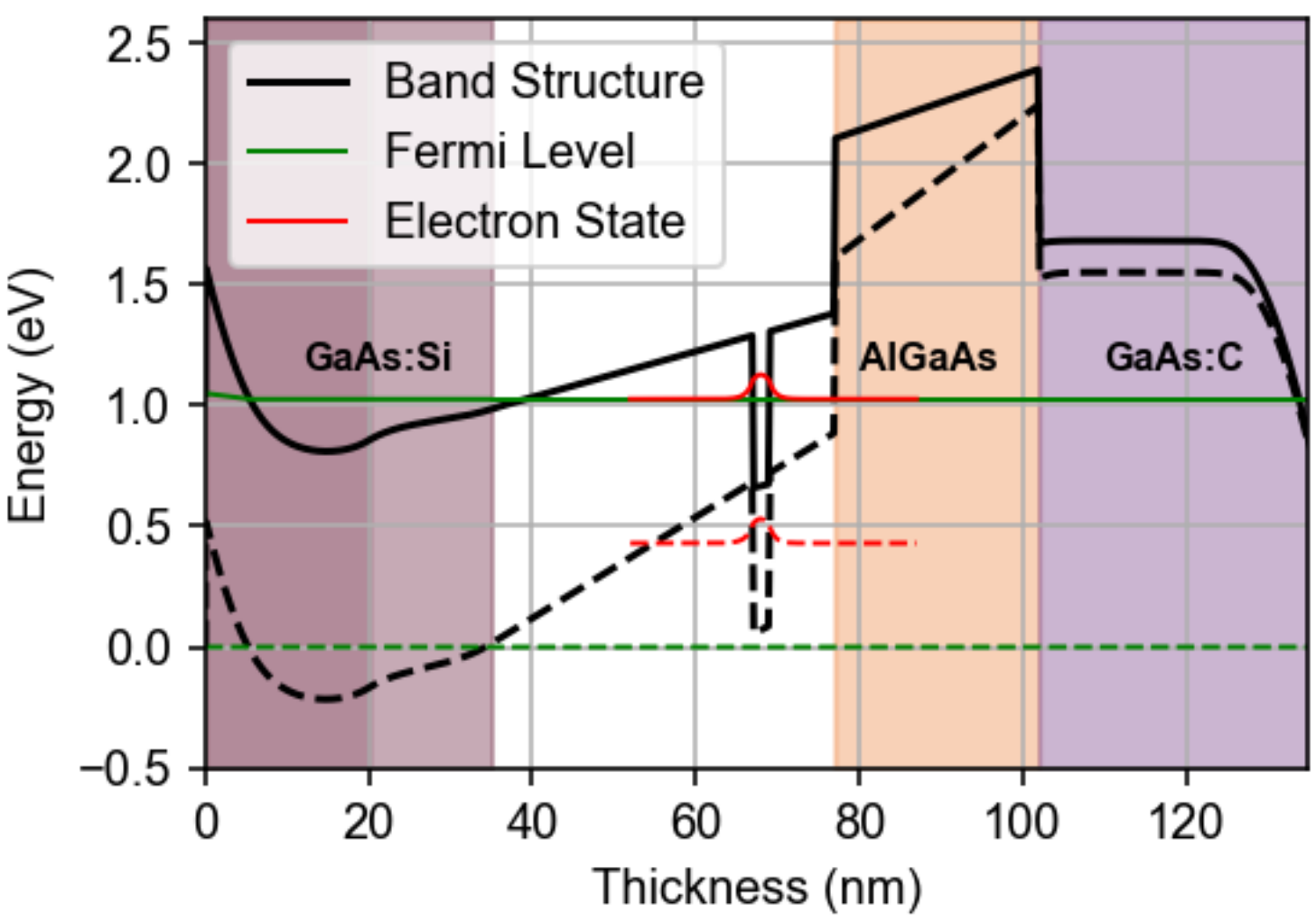

*Figure 1: Simulation of the conduction band of the device at T = 5 K, plotted at zero bias voltage $V_b$ = 0 (dashed lines) and forward bias voltage $V_b$ = 1.4 V (solid lines). The Fermi level for the n-doped areas (green) have been extended through the whole structure as a guide for the eye. At zero bias, the energy states in the QD are positioned well above the Fermi level of the electron reservoir. By applying a bias voltage, the band structure is flattened. In the simulation, at $V_b$ = 1.1 V the lowest energy states in the QD (red line) reach resonance with the Fermi level at the n-doped contact.*

The heterostructure was grown with a solid-source III-V MBE machine on an n-doped 3-inch GaAs (100) substrate. The dopants for the p-i-n diode are silicon (Si) and carbon (C) for the n- and p-type doping, respectively. After a 200 nm GaAs buffer layer and a 20 nm $Al_{0.85}Ga_{0.15}As$ sacrificial layer, the optically active membrane starts with 20 nm GaAs:Si with a nominal doping concentration of $1x10^{19}$ $cm^{-3}$, followed by 15 nm of GaAs:Si with $3x10^{18}$ $cm^{-3}$ and a layer of 30 nm of intrinsic GaAs.

The InAs QDs were grown using the Stranski-Krastanov growth mode under an arsenic pressure of $8x10^{-6}$ Torr. The QD growth sequence involved the nominal deposition of 1.6 monolayers (ML) of InAs at a substrate temperature of 510 °C, followed by a 15 s growth interruption, and a second deposition of 0.7 ML of InAs for a total of 2.1 ML. To target a 920 nm emission wavelength, the QDs are capped using partial capping and annealing [28] by overgrowing them with 3 nm of GaAs before increasing the substrate temperature to 580 °C. Afterwards, the QD-containing layer was overgrown with an additional 5 nm of i-GaAs, followed by a 25 nm $Al_{0.85}Ga_{0.15}As$ barrier and finally a layer of 35 nm GaAs:C with a nominal doping concentration of $1x10^{19}$ $cm^{-3}$.

## Device fabrication

Following epitaxial growth, a 5 x 7 $mm^2$ sample piece with a low spectral QD density was selected. Based on hyperspectral imaging data, the spectral density of QDs in the 920±10 nm range was estimated to be around $5x10^{-6}$ $cm^{-2}$. To form the backside mirror, a 330 nm $Al_2O_3$ spacer layer was deposited by sputtering, followed by a 100 nm gold layer. The gold was deposited using a lithographically patterned resist mask to keep the future electrical contact areas uncovered.

The structure was then flip-chip bonded onto a GaAs carrier substrate using a 2 µm layer of benzocyclobutene (BCB) as an adhesive. The sample was cured for 2 hours at 150 °C under an applied pressure of approximately 40 kPa. Subsequently, the original GaAs substrate and buffer layer were removed through a combination of mechanical lapping and wet chemical etching using a mixture of citric acid and hydrogen peroxide. The $Al_{0.85}Ga_{0.15}As$ sacrificial layer was removed with 5% hydrofluoric acid, revealing the p-i-n doped membrane.

Adding the electrical contacts required several successive lithography steps. High-resolution hyperspectral imaging markers were defined via electron-beam lithography (EBL), followed by gold evaporation and lift-off. Then the membrane was patterned into six separate mesas using an optical mask and dry etching. Within each mesa, a central strip was removed by selective wet etching down to the p-doped GaAs layer, effectively dividing the mesa into two regions with separate n-doped contacts. This configuration allows each mesa to form two independent diodes sharing a common p-type contact. To prevent electrical shorts, especially at the mesa sidewalls, the sample surface excluding the mesas was passivated with $HfO_2$. Finally, gold bonding pads and electrical connection paths to the p- and n-type layers were evaporated onto the sample. The resulting schematic cross-section of the processed structure is shown in Figure 2a.

In the next step, the positions of single QDs with a suitable emission wavelength were identified using the hyperspectral imaging technique described by Buchinger et al. [29]. The CBGs were then deterministically placed around pre-selected QDs using high-resolution EBL with a PMMA resist, followed by reactive ion etching. Figure 2b shows a microscope image of two diode regions (bottom left in the image) separated by the gold p-side contact area before CBG fabrication, while Figure 2c shows a zoomed in picture after bullseye placement, with the n-type gold contact visible on the left. Each bullseye has been individually placed around a single QD. Alongside the CBG fabrication, the diode area was trimmed to reduce its overall footprint.

Because the Purcell factor is highly sensitive to the CBG geometry relative to the emission wavelength, the radius of each CBG center disk was individually optimized to match the specific

QD emission. To maintain electrical connectivity with the central disk without sacrificing optical performance, we employed a labyrinth CBG design with fourfold symmetry and ~115 nm wide bridges [25]. Figure 2d shows an SEM image of a complete labyrinth bullseye CBG. The smooth surface and well-defined ridges indicate high fabrication quality.

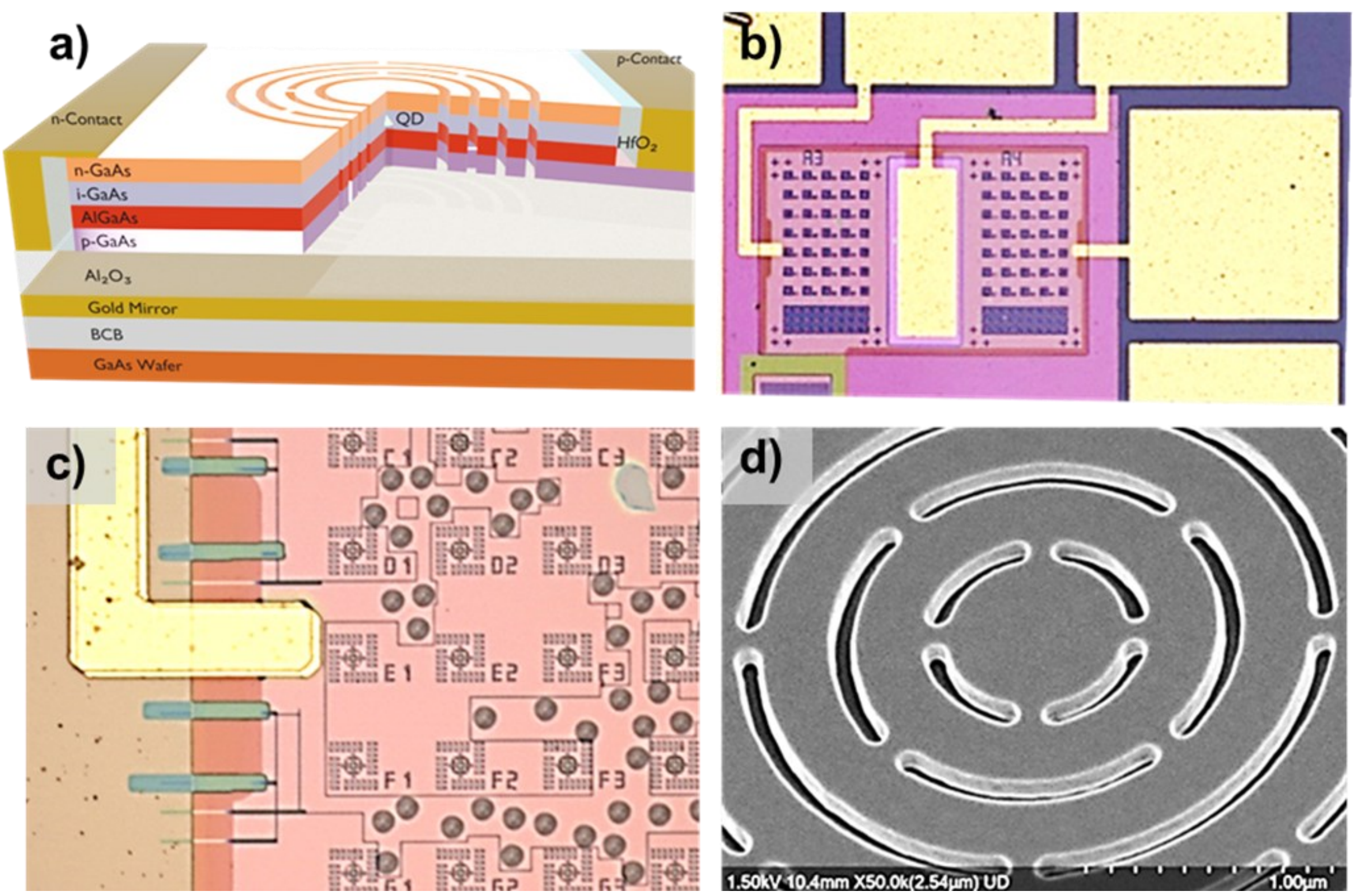

*Figure 2:a) Schematic cross section of the device, showing membrane with the CBG, the $Al_2O_3$ spacer layers, bottom gold mirror, glued with BCB to a GaAs wafer. The sidewalls between membrane and gold contacts are filled with $HfO_2$ to prevent electrical shorts b) Microscope image of two measurement fields, showing the golden electrical pathways leading to the bonding pads (two sides: n-type, middle: p-type). c) Microscope image zoomed into a measurement field on the n-type contact side d) SEM image of processed bullseye, showing clean surfaces and well-defined ridges.*

## Device Characterization

The sample was wire-bonded with gold wires and cooled to 2 K in a closed-cycle cryostat. Current-voltage (I-V) measurements exhibit characteristic diode behavior (Figure 3a). Above-bandgap micro-photoluminescence (µPL) measurements of a deterministically placed CBG reveal a cavity mode centered at 910 nm with a full width at half maximum (FWHM) of 4.6 nm (b). The mode is spectrally well-aligned with the QD emission and possesses sufficient bandwidth to encompass both X and $X^-$ transitions, even when accounting for the bias-dependent Stark shift of around 1.2 meV (0.4 nm).

Polarization-resolved µPL of the cavity mode shows a minimal degree of linear device polarization of less than 10%. This confirms the high degree of symmetry of the fabricated CBG and the precise alignment of the cavity relative to the QD [29,30]. Photoluminescence excitation (PLE) spectroscopy reveals several resonant excitation channels as a function of the excitation wavelength (Figure 3c), including quasi-resonant excitation via the wetting layer at a laser wavelength of 888 nm and p-shell excitation at 897 nm.

d displays a color-coded map of µPL intensity versus bias voltage (PL-V) under above-bandgap excitation. The gate voltage is applied on the n-doped side as a DC bias. The PL-V spectrum initially exhibits a distinct blueshift around 2.9 V. We attribute this to an unintentional lateral electric field component that competes with the vertical quantum-confined Stark effect (QCSE) [31]. One argument is that our initial simulations (see Figure 1) suggest that we should observe emissions at a bias voltage of 1.1 V. Furthermore, the voltage needed to observe emission increases for devices farther away from the p-contact. Both observations agree with a lateral field observation. The lateral field induces a spatial separation of the electron and hole wavefunctions, leading to an initial blueshift [32,33]. As the bias voltage increases further, the band structure is flattened and the conventional QCSE [34] dominates, resulting in a characteristic redshift.

To identify the charge states of the observed emission lines, we performed power-dependent and linear polarization-resolved µPL spectroscopy at several different bias voltages. Three distinct plateaus are visible in the PL-V spectrum (Figure 3d). First, we identified the lines ranging between 3 V and roughly 4.2 V, as neutral exciton (X), biexciton (XX) and negatively charged trion ($X^-$). XX and X exhibit a strong linear polarization dependency of the emission wavelength with a relative orientation offset of 45° in their polarization axis. The measured fine-structure splitting (FSS) of 63± 2 µeV for the XX and 56 ± 1.5 µeV for X are typical values for InAs QDs [35]. In contrast, $X^-$ shows almost no linear polarization dependency, as expected from its singlet ground state. The nature of the XX transition was further confirmed by observing a cascade to the X transition, measured in cross-correlation.

In the second plateau (4 V to 5 V), we identified two doubly negatively charged excitation lines ($X^{2-}$) alongside a negatively charged biexciton ($XX^-$). Finally, starting at approximately 4.8 V, three $X^{3-}$ emission lines appear. These assignments were validated through a combination of polarization-dependent intensity and wavelength measurements and the energy difference relative to the neutral exciton [36], as well as second-order photon cross-correlation measurements.

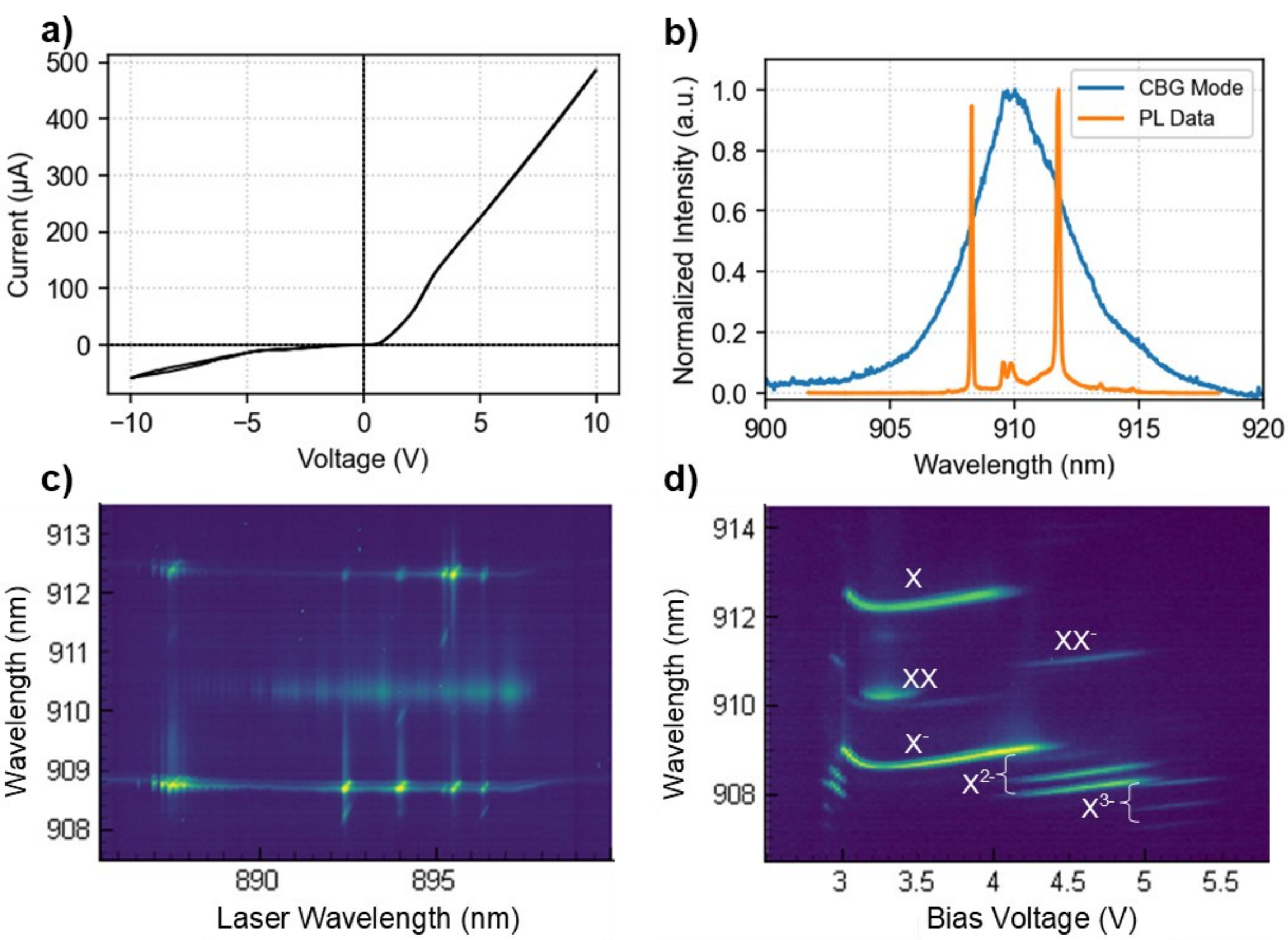

*Figure 3: a) Characteristic I-V line of the contacted p-i-n membrane exhibiting diode behavior. b) Above-bandgap PL spectrum showing the CBG mode at a bias voltage of 3.3 V. The CBG mode is centered around 910nm with a FWHM of 4.6 nm. c) PLE map showing resonant excitation channels as a function of the laser wavelength. d) Voltage dependent µPL (PL-V) map: three distinct charge plateaus are visible.*

**Optical Properties**

Quantum optical measurements were performed using a Ti:sapphire laser in pulsed picosecond mode (76 MHz) using quasi-resonant excitation (888 nm) via some unidentified higher exited state. The collection path contained a half-wave plate in cross-polarization orientation relative to the excitation path to filter out laser scattering. The $X^-$ line was further filtered using a transmission grating monochromator, passed through a 50/50 beam splitter, and measured with superconducting nanowire single-photon detectors (SNSPDs).

We calculate the photon extraction efficiency (PEE) of our device using the relation $PEE = \frac{n}{\eta_{Setup}*\eta_{Detector}*f}$, where n is the single photon emission rate, $\eta_{Setup}$ is the setup efficiency $\eta_{Detector}$ is the detector efficiency and f is the repetition rate. To determine the single-photon emission rate *n,* we obtained the saturation optical pump power under pulsed excitation for the trion line. A maximum count rate of n = 2.2 ± 0.1x$10^6$ $s^{-1}$ was detected at the SNSPDs. Combined with the measured setup collection efficiency $\eta_{Setup} = 0.05$, the nominal detector efficiency $\eta_{Detector}$ = 0.85, and the excitation laser frequency f = 76 MHz, we calculate a PEE of 68.1 ± 3.1%. However, further theoretical studies on improving the labyrinth CBG design predict that the PEE can reach above 90% [27]. In comparison, recent publications on ridge-based electrically contacted CBGs using a DBR as a backside mirror report measured PEE of up to 30.4 % and simulated PEE of 31.9 % and 56.5 % for optimized designs, respectively [17]. CBG antenna devices with a DBR bottom mirror achieve PEEs up to 24 % [37]. This significant increase of PEE demonstrated here represents an important improvement and is a step towards scalable QD-based photon sources for lager photonic cluster states [38] and quantum repeater building blocks [39].

To quantify the Purcell enhancement, the lifetime of the $X^-$ line was investigated in relation to the applied diode bias voltage Figure 4a shows the PL-V spectrum under pulsed quasi-resonant excitation (889nm) with a laser power of 13.5 µW. Compared to PL-V under above bandgap excitation (Figure 3d), emission lines other than $X^-$ are strongly subdued between 3.2 and 4.0 V, with the two $X^{2-}$ lines vanishing completely.

Time-resolved µPL measurements (Figure 4b) show an emission lifetime of $T_1$=44.3 ± 0.2 ps at a diode voltage of 3.6 V. To extract the correct value, the measurement data were fitted with a convoluted function consisting of the Gaussian instrument response function (IRF) of the SNSPD and an exponential decay function. Lifetime measurements of QDs in the planar membrane yielded a lifetime of 799 ± 30 ps for trion lines. Therefore, we calculate a Purcell factor of $F_P^{meas} =$ 18.0 ± 0.7 which is close to the theoretical value for our specific CBG layout of 22.

From the measured radiative lifetime $T_1$ we calculate a Fourier-transform limit $FTL = \frac{\hbar}{2\pi T_1}$ of 14.4 µeV for $X^-$. Comparing this value to the measured FWHM of the $X^-$ emission line of 64 µeV, we find a discrepancy by a factor of 4.4. One reason for this is that the quantum dots grown for this sample exhibit some innate line broadening, most likely due to unintentional impurities added during the MBE growth from the arsenic source. Other reasons may include charge noise [40] and other environmental factors.

Figure 4c shows several additional radiative lifetime measurements of $X^-$ at different diode voltages. A clear valley can be seen between 3.3 V and 3.8 V. This corresponds with a minimum in the FWHM of $X^-$ and an absence of $X^{2-}$ emission lines (Figure 4d).

Finally, we investigated the blinking behavior and the multi-photon contribution of our device. To this end, we performed time-resolved second order autocorrelation measurements in a Hanbury Brown and Twiss (HBT) configuration [41]. Figure 4e shows a photon autocorrelation histogram of the X⁻ line over a 1 µs time window at a diode bias of 3.6 V. No blinking behavior is observed, as the p-i-n diode suppresses this effect [42]. Figure 4f reveals very strong antibunching behavior, with the peak at 0 ns being entirely suppressed. No residual peak can be distinguished from the background noise in the dataset. An upper limit for the multi-photon contribution of 0.48% was estimated, by dividing the integrated area of the first peak by the integrated area around 0 ps with an integration window of two times the lifetime of the emission.

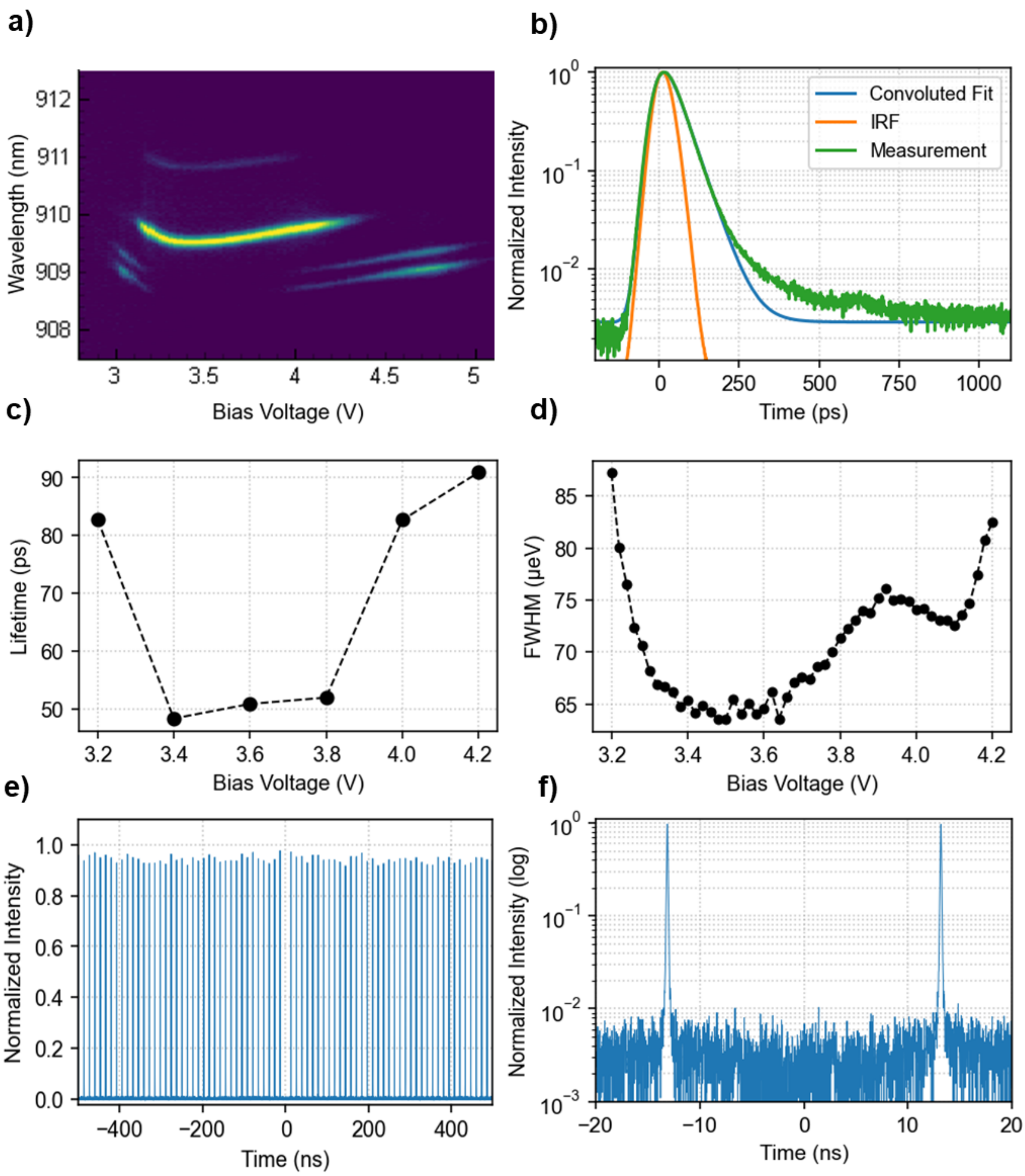

*Figure 4: Measurements under quasi-resonant excitation. a) PL-V focused on X⁻ emission. b) time resolved µPL measurement of X⁻ with a convoluted fit consisting of the SNSPD IRF and an exponential decay function. c) Radiative lifetime measurements of X⁻ at different diode voltages. d) FWHM of X⁻ emission line. e) Time-resolved second order autocorrelation measurement in Hanbury Brown and Twiss (HBT) configuration. f) Zoomed in around 0 ns with logarithmic y-scale.*

# Conclusion

In this paper, we demonstrated a charge-tunable single-photon source based on semiconductor QD in deterministically placed CBGs. We presented our experimental results, showcasing a blinking-free $X^-$ emission line under quasi-resonant excitation. $g^{(2)}(\tau)$ correlation measurements show a low multi-photon contribution for the $X^-$ line of less than 1.7% ± 1.5%

The labyrinth CBG design achieves a PEE of 68.1±3.1 % which is in good agreement with our simulated values of 58 % while simultaneously providing an electrical pathway to the QD. Further theoretical studies suggest that values of more than 90% are feasible, by optimizing the CBG parameters [27]. Compared to previously reported electrically contacted CBG devices, our device achieves an enhanced PEE, representing an important step towards scalable QD-based quantum photonic architectures. At the same time, our Purcell factor of $F_p$ = 18.0 ± 0.7 is close to the simulated value of 22 for our specific CBG layout.

Furthermore, we demonstrated charge control, with three charge plateaus visible in the PL-V spectrum, relating to $X^-$, $X^{2-}$ and $X^{3-}$, respectively. The presented design with an AlGaAs barrier on the p-doped side of the membrane favors negative charge states. By placing the AlGaAs barrier on the n-doped side, the device can be reconfigured to support positive charge states.

These results showcase the potential of this single-photon source for a wide range of quantum technology applications, where both precise charge control and excellent quantum optical properties are required.

# Declarations

## Competing Interests

The authors declare no competing interests.

## Availability of data and materials

The datasets used and/or analysed during the current study are available from the corresponding author on reasonable request.

## List of Abbreviations

| | |
|---|---|
| BCB | Benzocyclobutene |
| CBG | circular Bragg grating |
| EBL | electron-beam lithography |
| FTL | Fourier transform limit |
| FWHM | full width at half maximum |
| IRF | instrument response function |
| I-V | current voltage |
| MBE | molecular beam epitaxy |
| μPL | micro-photoluminescence |
| PEE | photon extraction efficiency |
| PLE | photoluminescence over excitation wavelength |
| PLV | photoluminescence over voltage |
| QD | quantum dots |
| QCSE | quantum-confined Stark effect |
| QFDC | quantum frequency down conversion |
| SEM | scanning electron microscope |
| SNSP | superconducting nanowire single-photon detector |
| SPS | single-photon source |
| TGM | transmission grating monochromator |
| X | exciton |
| $X^-$ | negatively charged trion |
| XX | biexciton |

## Funding

The authors acknowledge the support of the state of Bavaria and the Federal Ministry of Research, Technology and Space (BMFTR) within Projects QR.X (FKZ: 16KISQ010) and QR.N (FKZ: 16KIS2209 and 16KIS2206). THL acknowledges support from the BMFTR within the project Qecs (FKZ: 13N16272).

## Authors' Contribution

P.G. and Q.B. contributed equally to the conception and writing of this article. P.G. supervised the project and was responsible for epitaxial growth. Q.B carried out theoretical analysis and numerical simulations and was responsible for device fabrication and deterministic device placement. P.G, Q.B., K.C.M., M.H., B.S.D., R.K. and H.K. performed experiments. S.K. fabricated the device with the help of Q.B. and I.B.. A.L. and D.R. grew initial samples for the study. S.H. and T.H.L. supervised the work, provided ideas and support and were responsible for funding. All authors supported the writing of the article.